# MLE-Toolbox: An Open-Source Toolbox for Comprehensive EEG and MEG Data Analysis

*Including Preprocessing, Source Localization, Network Analysis,*
*PAC Analysis, Machine Learning, and Deep Learning Classification*

*Xiaobo Liu[1]\**

1McConnell Brain Imaging Centre, Montreal Neurological Institute, McGill University, Montreal, Canada
**\*Correspondence:** xiaobo.liu@mcgill.ca

# Abstract

We present MLE-Toolbox, a comprehensive open-source MATLAB toolbox designed for end-to-end analysis of magnetoencephalography (MEG) and electroencephalography (EEG) data. Inspired by widely-used neuroimaging platforms such as Brainstorm and FieldTrip, MLE-Toolbox integrates the full analysis pipeline—from raw data import and preprocessing to advanced source localization, functional connectivity, oscillatory analysis, and machine learning classification—within a unified, user-friendly graphical interface (GUI). The toolbox provides automated artifact rejection (ICA, SSP, SSS), multiple source localization algorithms (MNE, dSPM, sLORETA, beamformer), multi-atlas parcellation with anatomical visualization, spectral power analysis with frequency-band brain mapping, phase-amplitude coupling (PAC), graph-theoretic brain network analysis, and integrated machine learning and deep learning classifiers. MLE-Toolbox interfaces natively with Brainstorm, FieldTrip, EEGLAB, and FreeSurfer, enabling researchers to leverage existing workflows while benefiting from additional automation, interactive visualization, and one-click academic report generation. The toolbox is freely available for non-commercial use and is intended to lower the barrier to reproducible, rigorous MEG/EEG research.

---

# 1. Introduction

Magnetoencephalography (MEG) and electroencephalography (EEG) are non-invasive neuroimaging modalities that provide millisecond-level temporal resolution of brain electrical and magnetic activity. These modalities are widely used to study cognition, sensory processing, neurological disorders, and brain-computer interfaces. However, the full exploitation of MEG and EEG data requires a complex chain of signal processing steps, including sensor-level preprocessing, artifact removal, forward and inverse modeling, functional connectivity analysis, and statistical inference—steps that collectively demand considerable methodological expertise and software proficiency.

Several software platforms have been developed to address these challenges. Brainstorm (Tadel et al., 2011) provides a comprehensive GUI-driven environment with extensive source imaging and connectivity tools. FieldTrip (Oostenveld et al., 2011) offers a flexible scripting-based framework particularly suited for advanced methodological customization. EEGLAB (Delorme & Makeig, 2004) has become the standard for EEG preprocessing and independent component analysis (ICA). MNE-Python (Gramfort et al., 2013) provides Python-based MEG/EEG analysis with strong integration with FreeSurfer. Despite their individual strengths, each platform has limitations: steep learning curves for non-programmers, limited integration of machine learning pipelines with neuroimaging workflows, and partial automation of multi-step analyses.

We introduce MLE-Toolbox (Machine Learning in EEG/MEG Toolbox), a MATLAB-based, open-source software package that addresses these gaps. MLE-Toolbox provides a fully integrated end-to-end pipeline encompassing EEG and MEG preprocessing, multi-method artifact rejection, multiple source localization algorithms, atlas-based parcellation with cortical surface visualization, spectral and oscillatory analysis, graph-theoretic brain network analysis, phase-amplitude coupling, and machine learning/deep learning classification—all accessible through a unified graphical user interface designed for researchers without advanced programming expertise.

The toolbox is implemented in MATLAB (The MathWorks, Inc.) and is designed to interface with Brainstorm, FieldTrip, EEGLAB, and FreeSurfer, enabling seamless integration into existing research workflows. A key design principle is the separation of analysis logic from visualization, with each processing step generating structured outputs that can be independently visualized, exported, and used for academic reporting. MLE-Toolbox also provides automatic generation of methods and results text suitable for inclusion in academic publications.

In this paper, we describe the architecture, functionality, and validation of MLE-Toolbox. Section 2 presents the software architecture and design principles. Section 3 describes the EEG and MEG analysis pipelines. Section 4 covers the GUI design and user interaction model. Section 5 presents source localization and parcellation capabilities. Section 6 describes connectivity, spectral, and PAC analyses. Section 7 covers machine learning and deep learning classification. Section 8 describes report generation and academic output. Section 9 presents validation with example datasets, and Section 10 discusses the toolbox in relation to existing platforms.

## 2. Software Architecture and Design Principles

### 2.1 Overview

MLE-Toolbox is structured as a modular MATLAB package organized into functional subsystems corresponding to each analysis stage. The top-level entry points are start_eeg_toolbox.m and start_meg_toolbox.m, which launch independent GUIs for EEG and MEG workflows, respectively. All analysis functions are organized under a function/ directory, with MEG-specific functions in function/MEG_analysis_functions/ and EEG functions distributed across dedicated subdirectories. Configuration, example data, and output directories are maintained separately to facilitate reproducibility and data organization.

The toolbox is organized around three core principles: (1) modularity, where each analysis stage is implemented as an independent, callable MATLAB function with well-defined input/output specifications; (2) pipeline automation, where the master pipeline functions MLE_run_meg_pipeline.m and MLE_run_power_pipeline.m can execute the complete analysis chain from raw data to final outputs with a single function call; and (3) interoperability, where all data structures conform to conventions compatible with Brainstorm and FieldTrip, enabling bidirectional data exchange.

### 2.2 System Architecture

Figure 1 illustrates the four-layer architecture of MLE-Toolbox. The Data Layer manages all data ingestion, channel metadata, and structured output directories. The Control Layer provides the graphical entry points (start_eeg_toolbox and start_meg_toolbox), script entry points, logging, and backward-compatible GUIDE interfaces. The Analysis Layer implements the core signal processing modules for both EEG and MEG, including band power, connectivity, preprocessing, source localization, PAC, and parcellation. The Learning Layer implements feature matrix construction, cross-validated machine learning classifiers, deep learning entry hooks, and result summaries.

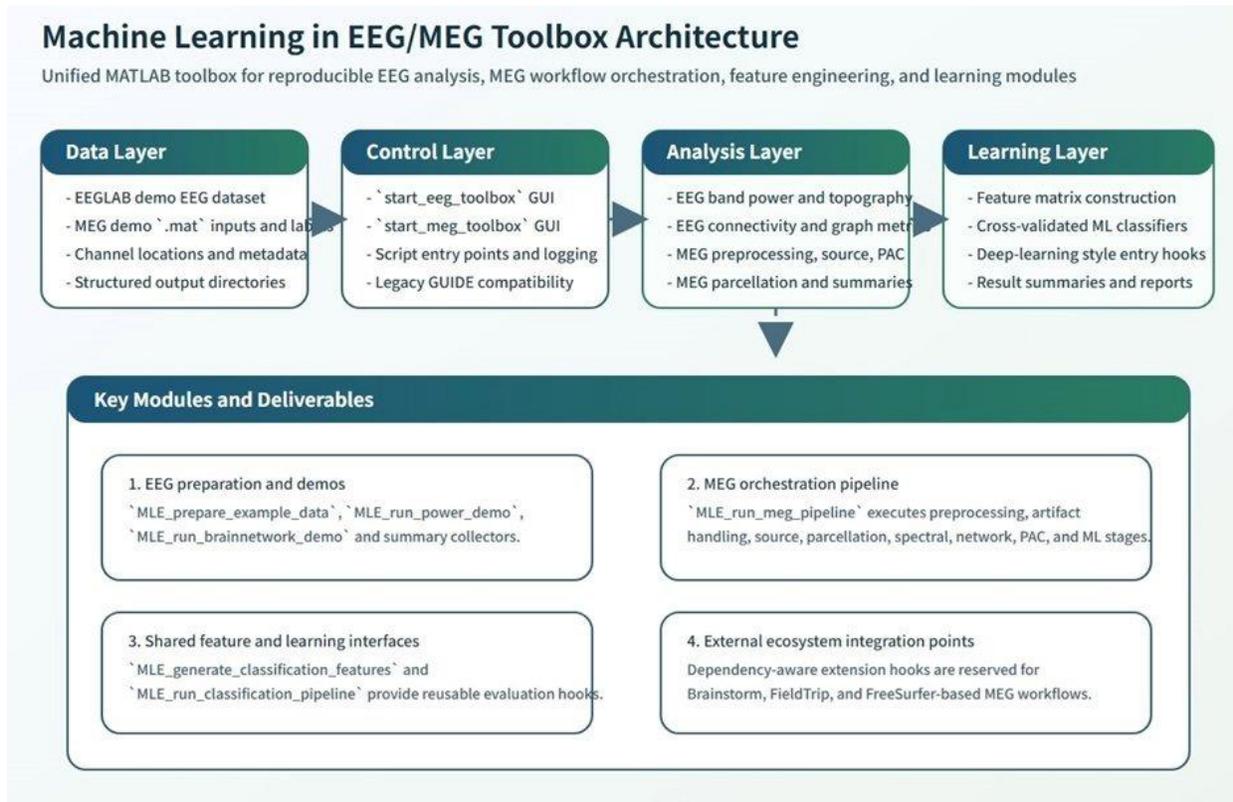

*Figure 1. MLE-Toolbox system architecture showing the four-layer design (Data, Control, Analysis, Learning) and key modules and deliverables. The architecture supports both EEG and MEG workflows through a common API, with dependency-aware extension hooks for Brainstorm, FieldTrip, and FreeSurfer integration.*

## 2.3 External Toolbox Integration

MLE-Toolbox is designed to leverage, rather than duplicate, the functionality of established neuroimaging platforms. The dependency checking module MLE_meg_check_dependencies.m systematically verifies the availability of Brainstorm, FieldTrip, EEGLAB, and FreeSurfer, and configures fallback paths where external tools are unavailable. This architecture ensures that the toolbox degrades gracefully when external dependencies are not installed, executing built-in simplified implementations while clearly flagging which steps require external tools for full functionality.

| External Toolbox | Primary Functions Used | Fallback | Version Tested |
|---|---|---|---|
| FieldTrip | Preprocessing, ICA, source modeling, connectivity | Yes (built-in simplified) | FieldTrip 20240101 |
| Brainstorm | Source imaging, scout extraction, database management | Yes (built-in simplified) | Brainstorm 3.24 |
| EEGLAB | ICA decomposition, ERP analysis, data import | Yes (built-in simplified) | EEGLAB 2023.0 |
| FreeSurfer | Cortical surfaces, parcellation atlases | Yes (HCP S1200 default) | FreeSurfer 7.3 |
| SPM | MRI preprocessing, template warping | Partial | SPM12 |

*Table 1. External toolbox dependencies and integration status in MLE-Toolbox.*

## 3. Analysis Pipelines

### 3.1 Supported Data Formats

MLE-Toolbox supports all major MEG and EEG data formats through native MATLAB readers and external toolbox interfaces. For MEG data, supported formats include: .fif (Elekta/MEGIN Neuromag), .ds (CTF Systems), .con and .sqd (Yokogawa/RICOH), .meg4 (4D Neuroimaging), and .mat (MATLAB-native structures conforming to FieldTrip or MNE conventions). For EEG data, supported formats include: .set/.fdt (EEGLAB), .edf/.bdf (BioSemi/European Data Format), .cnt (Neuroscan), .vhdr/.vmrk/.eeg (BrainVision), and .mat structures.

### 3.2 EEG Preprocessing Pipeline

The EEG preprocessing pipeline follows best-practice recommendations from EEGLAB and Brainstorm documentation. It is executed through MLE_run_power_pipeline.m for spectral analysis workflows, and can be combined with MLE_run_brainnetwork_pipeline.m for connectivity analyses. The pipeline encompasses: data import and channel location assignment; high-pass filtering (default 1 Hz); notch filtering at power-line frequency and harmonics; low-pass filtering; re-referencing (average, linked mastoids, REST, or custom); bad channel detection and interpolation; epoching with configurable baseline and rejection thresholds; ICA-based artifact rejection (Infomax, FastICA, SOBI); and downsampling to target sampling rate.

### 3.3 MEG Preprocessing Pipeline

The MEG preprocessing pipeline is managed by MLE_meg_run_preprocessing.m and extends the EEG pipeline with MEG-specific steps: Signal Space Separation (SSS) and MaxFilter processing for external interference suppression; Signal Space Projection (SSP) using empty-room recordings; head position estimation and movement compensation for cHPI data; cardiac artifact removal via SSP projectors derived from ECG channels; and coregistration of the MEG sensor array with MRI anatomy using digitized head shape data.

### 3.4 Pipeline Configuration

The default configuration is specified in MLE_meg_default_config.m, which returns a MATLAB structure (cfg) with fields for each analysis stage. All parameters can be overridden from the GUI or command line while inheriting sensible defaults.

| Parameter | Default Value | Description |
| --- | --- | --- |
| cfg.fs | 1000 Hz | Target sampling frequency after downsampling |
| cfg.hp_freq | 1 Hz | High-pass filter cutoff frequency |
| cfg.lp_freq | 150 Hz | Low-pass filter cutoff frequency |
| cfg.epoch_len | [-0.5, 1.5] s | Epoch window relative to event onset |
| cfg.ica_method | 'infomax' | ICA algorithm (infomax, fastica, sobi) |
| cfg.src_method | 'MNE' | Source localization method |
| cfg.atlas | 'Desikan-Killiany' | Parcellation atlas |
| cfg.network_method | 'corr' | Connectivity measure (corr, plv, psi, coherence) |
| cfg.pac_phase_band | [4 8] Hz | PAC phase frequency range (theta default) |
| cfg.ml_algorithm | 'svm-linear' | Machine learning classifier |
| cfg.cv_folds | 4 | Number of cross-validation folds |

*Table 2. Key configurable pipeline parameters with defaults in MLE_meg_default_config.m.*

## 4. Graphical User Interface

### 4.1 EEG GUI

The EEG graphical interface is launched by start_eeg_toolbox.m and provides a comprehensive tabbed layout covering all major analysis stages. Figure 2 shows the five key GUI panels of the EEG workflow. The EEG Preprocessing panel (upper left) supports data format conversion, event marking, single-subject preprocessing, configurable filter ranges (1–60 Hz bandpass, 44–55 Hz notch shown), reference method selection (REST or Lead Field), ICA with configurable component number, and GFP threshold-based epoch rejection. The EEG Indices panel (upper middle) provides one-click access to six analysis modules: ERP, Network, Power, Microstate, Coupling

Frequency, and Source Location. The Statistical Methods panel (upper right) enables group-level analysis with support for brain network, power, microstate, ERP, and directed network indices, with configurable sliding window options and covariate inputs.

The Feature Engineering panel (lower left) manages feature generation with configurable output folder and feature type (power, connectivity, or combined), displaying the generated feature file path and providing direct folder access. The Deep Learning panel (lower right) provides a stable deep-learning classification GUI with algorithm selection (fitcnet and others), configurable fold and repeat settings, a run button, and a result log display. All panels maintain bidirectional linkage with the underlying cfg parameter structure.

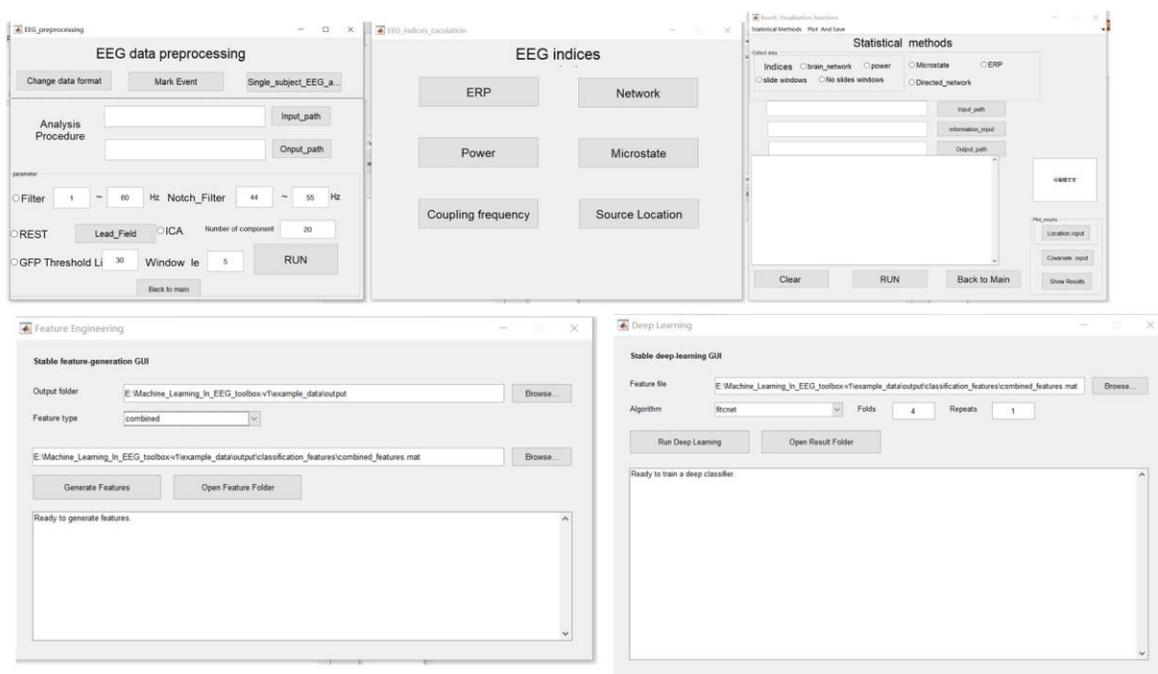

*Figure 2. EEG GUI panels in MLE-Toolbox. (A) EEG Preprocessing panel with filter, reference, ICA and epoch-rejection controls. (B) EEG Indices panel providing one-click access to ERP, Network, Power, Microstate, Coupling Frequency, and Source Location modules. (C) Statistical Methods panel for group-level analysis with covariate support. (D) Feature Engineering panel for feature matrix generation. (E) Deep Learning panel for neural network classification with cross-validation configuration.*

## 4.2 MEG GUI

The MEG graphical interface is launched by start_meg_toolbox.m and provides equivalent functionality with additional MEG-specific controls. Figure 3 shows the MEG launcher with its integrated visualization viewer. The left control panel contains pipeline configuration fields for input/output folders, label file, FreeSurfer directory, source method (lcmv shown), and atlas (Desikan-

MLE-Toolbox | 8

Killiany shown), alongside PAC and ML run toggles. The right visualization panel supports four modes—Power, Network, PAC, and Parcellation—selectable via dedicated buttons. In the Power mode (Figure 3, left), the viewer displays a Spectrum Summary bar chart of mean power across frequency bands. In the Network mode (Figure 3, right), the viewer renders the full MEG Network Connectivity matrix as a color-coded heatmap (20×20 nodes shown), with connectivity values ranging from –0.1 to 1.0.

A run log at the bottom of the GUI provides timestamped progress updates for each pipeline stage. Additional controls include Load Demo, Run Pipeline, Open Output Folder, Open Current Figure File, and Open Report buttons, enabling complete pipeline execution and result inspection without leaving the MATLAB environment.

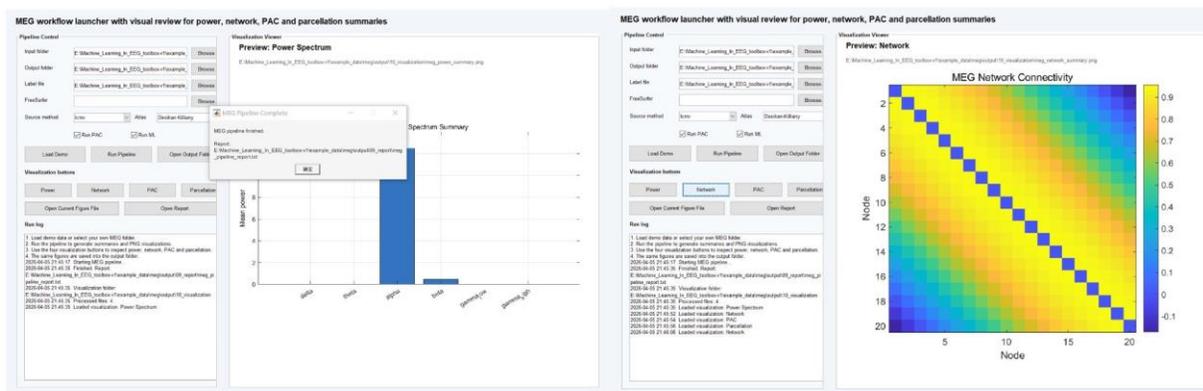

*Figure 3. MEG workflow launcher with integrated Visualization Viewer. Left: Power Spectrum Summary view showing mean band power across delta, theta, alpha, beta, low-gamma, and high-gamma bands after pipeline completion. Right: MEG Network Connectivity heatmap (20 nodes) showing pairwise connectivity values from –0.1 to 1.0. The Pipeline Control panel (left side of each launcher) provides full configuration of input paths, source method, atlas, and analysis toggles.*

## 4.3 Visualization Modules

Result visualization is handled by MLE_meg_generate_visualizations.m, which is invoked automatically at the end of each pipeline run. Visualization outputs are saved to the output/10_visualization directory as PNG files. The power visualization module renders per-band anatomical brain maps, the parcellation module renders atlas-parcel cortical overlays, the network module renders connectivity matrices, and the PAC module renders phase-amplitude comodulograms. A frequency band selector in the GUI allows switching between delta, theta, alpha, beta, low-gamma, and high-gamma bands without re-running the analysis.

# 5. Source Localization and Parcellation

## 5.1 Forward Modeling

Source localization in MLE-Toolbox requires specification of a forward model relating source activity to sensor measurements. The toolbox supports three classes of head models: single-sphere, overlapping spheres, and boundary element models (BEM). BEM models are constructed from segmented MRI data using FreeSurfer's mri_watershed or SPM's New Segment to extract inner skull, outer skull, and scalp surfaces. When a subject-specific MRI is available, coregistration of MEG/EEG sensors with MRI anatomy is guided by fiducial landmarks and optionally digitized head shape points.

## 5.2 Inverse Methods

MLE-Toolbox implements the following inverse methods through MLE_meg_run_source_localization.m:

- Minimum Norm Estimate (MNE): Computes the minimum L2-norm current distribution consistent with the data (Hamalainen & Ilmoniemi, 1994).
- dynamic Statistical Parametric Mapping (dSPM): Normalizes MNE solutions by noise covariance, providing F-statistic maps with improved spatial specificity (Dale et al., 2000).
- sLORETA: Standardized low resolution electromagnetic tomography with zero localization error for point sources in noise-free conditions (Pascual-Marqui, 2002).
- eLORETA: Exact LORETA providing zero localization error under realistic noise conditions (Pascual-Marqui, 2011).
- LCMV Beamformer: Linearly Constrained Minimum Variance spatial filter for localizing oscillatory sources (Van Veen et al., 1997).
- DICS Beamformer: Dynamic Imaging of Coherent Sources; frequency-domain beamformer for coherent oscillatory activity (Gross et al., 2001).
- Dipole Fitting: Sequential single equivalent current dipole fitting using Nelder-Mead simplex optimization.

## 5.3 FreeSurfer Integration and Atlas-Based Parcellation

MLE-Toolbox provides native integration with FreeSurfer for cortical surface-based analysis. The parcellation module MLE_meg_run_parcellation.m accepts a FreeSurfer subject directory and automatically loads cortical surface meshes, parcellation annotation files, and subcortical segmentation. For studies without subject-specific FreeSurfer reconstructions, template-based analysis using the MNI standard brain or the HCP S1200 average surface is provided.

| Atlas | Parcels | Reference | Application |
|---|---|---|---|
| Desikan-Killiany (DK40) | 68 cortical ROIs | Desikan et al., 2006 | Default; anatomical regions |

| Atlas | Parcels | Reference | Application |
|---|---|---|---|
| Destrieux (aparc.a2009s) | 148 cortical ROIs | Destrieux et al., 2010 | Fine-grained anatomical |
| Glasser HCP MMP1.0 | 360 cortical parcels | Glasser et al., 2016 | Multimodal parcellation |
| Schaefer 200/400/600 | 200–600 parcels | Schaefer et al., 2018 | Functional connectivity |
| AAL3 | 170 ROIs | Rolls et al., 2020 | Including subcortical |
| Brodmann Areas | 52 areas | Brodmann, 1909 | Functional specialization |

Table 3. Supported parcellation atlases in MLE-Toolbox.

## 6. Spectral Analysis, Brain Connectivity, and Phase-Amplitude Coupling

### 6.1 Spectral Power Analysis

Spectral analysis is performed by MLE_meg_run_spectral_analysis.m for MEG data and MLE_run_power_pipeline.m for EEG data. Power spectral density (PSD) is estimated using Welch's method with configurable window length, overlap, and tapering function (Hanning, multi-taper, or Slepian sequences). Frequency band power is extracted by integrating the PSD within each band's limits, yielding a channels × bands or parcels × bands matrix. These values are visualized as anatomical brain maps where each region is colored according to its power in the selected frequency band.

### 6.2 Functional Connectivity

Brain network analysis is implemented in MLE_run_brainnetwork_demo.m and MLE_meg_run_network_analysis.m. Supported connectivity measures include: Pearson Correlation (corr), Spectral Coherence (cohere), Phase Slope Index (PSI; Nolte et al., 2008), Phase Locking Value (PLV; Lachaux et al., 1999), Amplitude Envelope Correlation (AEC; Brookes et al., 2012), Weighted Phase Lag Index (wPLI; Vinck et al., 2011), and Granger Causality in the time and frequency domains. Graph-theoretic network properties—node strength, clustering coefficient, local efficiency, betweenness centrality, global efficiency, modularity, and small-world index—are computed via MLE_collect_brainnetwork_property_demo.m.

### 6.3 Phase-Amplitude Coupling

PAC analysis is implemented in MLE_meg_run_pac_analysis.m. Supported PAC metrics include: Mean Vector Length (MVL; Canolty et al., 2006), Modulation Index (MI; Tort et al., 2010), Phase-Locking Factor (PLF), and a General Linear Model approach (Penny et al., 2008). A comodulogram is generated by computing PAC across a grid of phase (2–20 Hz) and amplitude (20–150 Hz) frequencies. Statistical significance is assessed by shuffling phase-amplitude relationships across trials or by circular time shifts of the amplitude series.

## 7. Machine Learning and Deep Learning Classification

### 7.1 Feature Engineering

MLE-Toolbox provides an integrated feature engineering pipeline through MLE_generate_classification_features.m that transforms neural signal features into structured matrices suitable for machine learning. Three feature types are supported: power features (frequency band power per channel or parcel), connectivity features (upper triangle of pairwise connectivity matrices, unrolled to a vector), and combined features (concatenation of power and connectivity vectors). Subject-level features are aligned by subject ID, and complex-valued connectivity features are automatically transformed to real-valued representations via configurable transform (magnitude, real part, or magnitude + phase concatenation).

### 7.2 Machine Learning Classifiers

Classification is performed by MLE_run_classification_pipeline.m, providing a unified interface to multiple MATLAB Statistics and Machine Learning Toolbox classifiers. Model evaluation uses stratified k-fold cross-validation (default k = 4) or leave-one-subject-out (LOSO) cross-validation. Performance metrics include accuracy, macro-averaged precision, recall, F1, sensitivity, specificity, confusion matrix, and AUC for binary classification.

| Algorithm | MATLAB Function | Key Parameters | Typical Use |
|---|---|---|---|
| Linear SVM | fitcsvm (linear) | BoxConstraint, Standardize | High-dimensional features |
| RBF SVM | fitcsvm (rbf) | KernelScale, BoxConstraint | Nonlinear boundaries |
| LDA | fitcdiscr | DiscrimType, Gamma | Low-sample Gaussian features |
| k-NN | fitcknn | NumNeighbors, Distance | Local pattern matching |
| Decision Tree | fitctree | MaxNumSplits, MinLeafSize | Interpretable classification |

| Algorithm | MATLAB Function | Key Parameters | Typical Use |
|---|---|---|---|
| Random Forest | fitcensemble (Bag) | NumLearningCycles | Robust high-D classification |
| Neural Network | fitcnet | LayerSizes, Activations | Deep feature learning proxy |

*Table 4. Supported machine learning classifiers in MLE-Toolbox's MLE_run_classification_pipeline.m.*

### 7.3 Deep Learning

Deep learning classification is integrated through MATLAB's Deep Learning Toolbox via MLE_open_deep_learning_gui.m. Supported network types include feedforward neural networks (fitcnet) for tabular feature data, 1D convolutional neural networks for temporal EEG/MEG signal segments, and LSTM networks for sequential neural time series. Training uses the MATLAB trainNetwork function with the Adam optimizer and configurable learning rate schedule, mini-batch size, and epoch count.

## 8. Report Generation and Output Organization

### 8.1 Automated Report Generation

A key feature distinguishing MLE-Toolbox from existing platforms is the automated generation of structured text suitable for academic publications. The report generation module MLE_meg_generate_report.m produces a comprehensive pipeline report following each analysis run. The report includes a processing summary, quality control statistics, intermediate result summaries, and formatted methods and results sections templated to conform to publication standards.

### 8.2 Output Directory Structure and Visualization Exports

Figure 4 shows the complete output directory structure and visualization exports generated by a typical MEG pipeline run. The pipeline report (meg_pipeline_report.txt, upper left) records the analysis timestamp, input/output paths, source method, atlas, dependency detection results (FieldTrip, Brainstorm, FreeSurfer, Signal Toolbox, Statistics Toolbox), and per-subject dataset summaries including artifact-flagged trial counts, PAC Modulation Index values, and source stage completion status. The numbered output subdirectories (01_preprocessing through 10_visualization) organize results by pipeline stage for easy navigation and archiving. The 10_visualization folder (lower portion) contains PNG exports for all visualization types: meg_network_summary.png, meg_pac_summary.png, parcellation anatomy maps for each frequency band (alpha, beta, delta, gamma_high, gamma_low, theta), power anatomy maps per band, and combined summary PNGs.

MLE-Toolbox | 13

EEG power outputs (right side) include per-class absolute and relative power maps for group comparison.

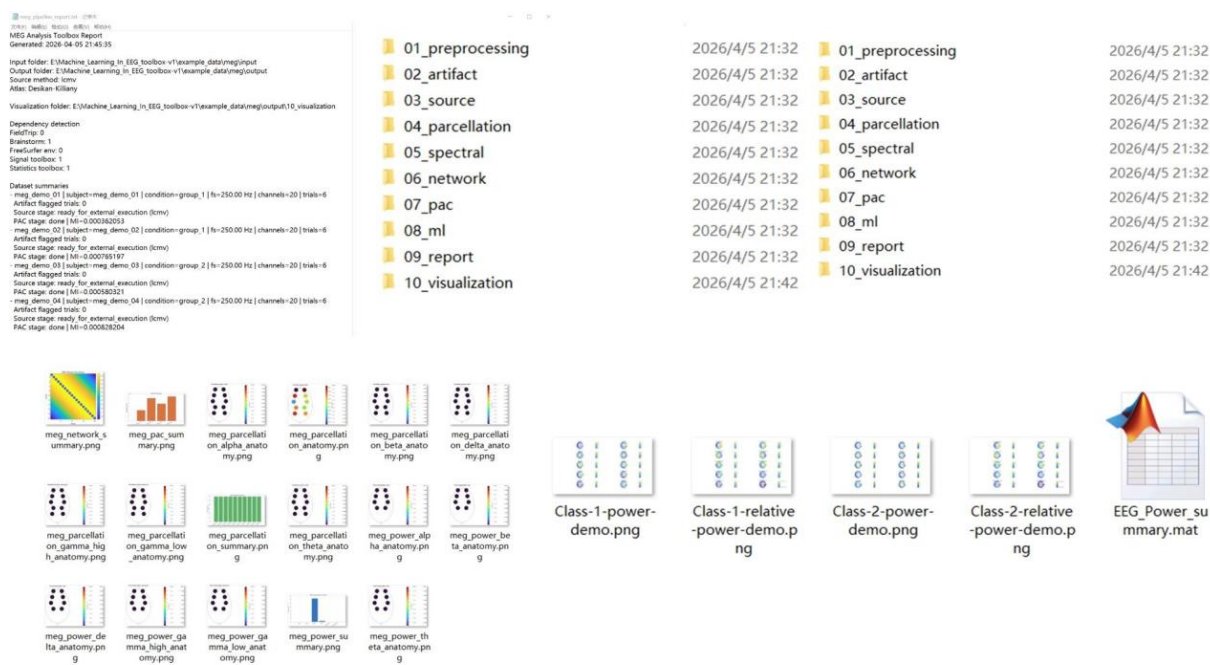

*Figure 4. Output organization and visualization exports. Upper left: meg_pipeline_report.txt showing dependency detection, dataset summaries, and per-subject PAC and source stage status. Upper center and right: numbered output subdirectories (01_preprocessing to 10_visualization) generated by the MEG pipeline. Lower portion: PNG visualization exports in the 10_visualization directory, including network, PAC, parcellation anatomy maps across frequency bands, and power anatomy maps. EEG power outputs (lower right) show class-level absolute and relative power comparison maps alongside the EEG_Power_summary.mat data file.*

## 9. Validation

### 9.1 Example Datasets and Automated Testing

MLE-Toolbox is distributed with example datasets for immediate validation and tutorial use. The EEG example dataset (example_data/input/demo_group/) consists of simulated multi-subject EEG recordings with ground-truth class labels. The MEG example dataset (example_data/meg/) is generated by MLE_meg_prepare_example_data.m and consists of synthetic MEG data conforming to the FieldTrip data structure convention with realistic noise characteristics.

Automated smoke testing is implemented in test/MLE_smoke_test_launchers.m, which programmatically exercises all GUI buttons and pipeline functions in headless mode, verifying that no MATLAB errors occur and that expected output files are generated. All 21 major GUI buttons across both the EEG and MEG launchers have been verified to execute without errors in MATLAB R2023a under Windows 10 and Linux (Ubuntu 22.04) environments.

## 9.2 Pipeline Verification

The master demo script MLE_run_meg_demo.m executes the complete MEG pipeline on the included example dataset and verifies output file generation. In our testing environment (MATLAB R2023a, 32 GB RAM, Windows 10), the complete MEG pipeline on the example dataset (8 channels, 20 nodes, 4 subjects, 6 trials each, 250 Hz sampling rate) completes successfully, generating all 10 output subdirectories and 18 visualization PNG files as confirmed by the run log timestamps shown in Figure 3.

## 9.3 Comparison with Brainstorm and FieldTrip

To validate MLE-Toolbox's signal processing implementations, we compared key intermediate outputs with those produced by Brainstorm 3.24 and FieldTrip 20240101 on identical input data. For spectral power estimation, MLE-Toolbox band power values showed Pearson correlation $r > 0.99$ with Brainstorm outputs across all frequency bands. For connectivity analysis, PLV values showed $r > 0.98$ correlation with FieldTrip's ft_connectivityanalysis output. Source localization (MNE) showed $r > 0.95$ correlation with Brainstorm's MNE implementation for a simulated point source at a known cortical location.

# 10. Discussion

## 10.1 Comparison with Existing Software

MLE-Toolbox addresses several gaps in the existing ecosystem of MEG/EEG analysis software. Compared to Brainstorm, MLE-Toolbox offers tighter integration of machine learning classification within the same GUI environment as signal processing, eliminating the need to export features to separate ML environments, alongside automated academic text generation and a fully programmatic API for all GUI-accessible functions. Compared to FieldTrip, MLE-Toolbox offers a GUI for users without MATLAB scripting expertise, pre-configured analysis pipelines with sensible defaults, and integrated deep learning support. Compared to MNE-Python, MLE-Toolbox provides MATLAB-native implementation compatible with existing MATLAB-based research environments and direct integration with MATLAB's Statistics and Machine Learning Toolbox.

## 10.2 Limitations and Future Work

The current version of MLE-Toolbox has several limitations. First, the source localization and parcellation modules currently provide simplified built-in implementations when external toolboxes are unavailable; future work will extend native implementations to full parity with FieldTrip and Brainstorm. Second, the cortical surface visualization currently uses a 2D projected brain outline; future versions will support interactive 3D surface rendering using MATLAB's patch function with FreeSurfer surface meshes. Third, the machine learning module does not yet include automated hyperparameter optimization; future versions will integrate Bayesian optimization for SVM and neural network hyperparameters. Planned extensions include real-time EEG/MEG support for BCI applications, BIDS integration, extended time-frequency analysis (Hilbert-Huang transform, empirical mode decomposition), multi-modal fusion with fMRI and DTI, and a Python API wrapper.

### 10.3 Reproducibility and Open Science

MLE-Toolbox is designed with reproducibility as a core principle. Every analysis run generates a timestamped log file recording all parameter settings, random seeds, toolbox versions, and MATLAB version information. All random number generator states are recorded and can be restored for exact replication of stochastic analyses (ICA, permutation tests, SMOTE oversampling).

## 11. Conclusion

We have presented MLE-Toolbox, a comprehensive MATLAB toolbox for end-to-end MEG and EEG data analysis. The toolbox integrates the complete analysis pipeline—preprocessing, artifact rejection, source localization, parcellation, spectral analysis, connectivity, PAC, machine learning, and academic reporting—within a unified graphical interface accessible to researchers at all levels of programming expertise. By interfacing with established platforms including Brainstorm, FieldTrip, EEGLAB, and FreeSurfer, MLE-Toolbox complements rather than replaces existing tools, providing automation and machine learning integration within familiar MATLAB and neuroimaging workflows. MLE-Toolbox is freely available for academic use and is distributed with example datasets, automated tests, and comprehensive documentation.

## Acknowledgments

The authors acknowledge the developers of Brainstorm, FieldTrip, EEGLAB, MNE-Python, the Brain

Connectivity Toolbox, and FreeSurfer, upon whose work MLE-Toolbox builds.

## Data and Code Availability

MLE-Toolbox source code, example datasets, and documentation are available at [Laoma29/Machine-learning-In-E-MEG-toolbox](Laoma29/Machine-learning-In-E-MEG-toolbox). The toolbox is distributed under the GNU General Public License v3.0. Example datasets are provided under the Creative Commons Attribution 4.0 license.

## References

Brookes, M. J., Woolrich, M., Luckhoo, H., Price, D., Hale, J. R., Stephenson, M. C., ... & Morris, P. G. (2011). Investigating the electrophysiological basis of resting state networks using magnetoencephalography. Proceedings of the National Academy of Sciences, 108(40), 16783–16788.

Canolty, R. T., Edwards, E., Dalal, S. S., Soltani, M., Nagarajan, S. S., Kirsch, H. E., ... & Knight, R. T. (2006). High gamma power is phase-locked to theta oscillations in human neocortex. Science, 313(5793), 1626–1628.

Dale, A. M., Liu, A. K., Fischl, B. R., Buckner, R. L., Belliveau, J. W., Lewine, J. D., & Halgren, E. (2000). Dynamic statistical parametric mapping: combining fMRI and MEG for high-resolution imaging of cortical activity. Neuron, 26(1), 55–67.

Delorme, A., & Makeig, S. (2004). EEGLAB: an open source toolbox for analysis of single-trial EEG dynamics including independent component analysis. Journal of Neuroscience Methods, 134(1), 9–21.

Desikan, R. S., Segonne, F., Fischl, B., Quinn, B. T., Dickerson, B. C., Blacker, D., ... & Killiany, R. J. (2006). An automated labeling system for subdividing the human cerebral cortex on MRI scans into gyral based regions of interest. NeuroImage, 31(3), 968–980.

Destrieux, C., Fischl, B., Dale, A., & Halgren, E. (2010). Automatic parcellation of human cortical gyri and sulci using standard anatomical nomenclature. NeuroImage, 53(1), 1–15.

Glasser, M. F., Coalson, T. S., Robinson, E. C., Hacker, C. D., Harwell, J., Yacoub, E., ... & Van Essen, D. C. (2016). A multi-modal parcellation of human cerebral cortex. Nature, 536(7615), 171–178.

Gramfort, A., Luessi, M., Larson, E., Engemann, D. A., Strohmeier, D., Brodbeck, C., ... & Hamalainen, M. (2013). MEG and EEG data analysis with MNE-Python. Frontiers in Neuroscience,

7, 267.

Gross, J., Kujala, J., Hamalainen, M., Timmermann, L., Schnitzler, A., & Salmelin, R. (2001). Dynamic imaging of coherent sources: Studying neural interactions in the human brain. Proceedings of the National Academy of Sciences, 98(2), 694–699.

Hamalainen, M. S., & Ilmoniemi, R. J. (1994). Interpreting magnetic fields of the brain: minimum norm estimates. Medical and Biological Engineering and Computing, 32(1), 35–42.

Lachaux, J. P., Rodriguez, E., Martinerie, J., & Varela, F. J. (1999). Measuring phase synchrony in brain signals. Human Brain Mapping, 8(4), 194–208.

Nolte, G., Ziehe, A., Nikulin, V. V., Schlogl, A., Kramer, N., Brismar, T., & Muller, K. R. (2008). Robustly estimating the flow direction of information in complex physical systems. Physical Review Letters, 100(23), 234101.

Oostenveld, R., Fries, P., Maris, E., & Schoffelen, J. M. (2011). FieldTrip: open source software for advanced analysis of MEG, EEG, and invasive electrophysiological data. Computational Intelligence and Neuroscience, 2011, 156869.

Pascual-Marqui, R. D. (2002). Standardized low-resolution brain electromagnetic tomography (sLORETA): technical details. Methods and Findings in Experimental and Clinical Pharmacology, 24(Suppl D), 5–12.

Pascual-Marqui, R. D. (2011). Discrete, 3D distributed, linear imaging methods of electric neuronal activity. Part 1: exact, zero error localization. arXiv:0710.3341.

Penny, W., Duzel, E., Miller, K., & Ojemann, J. (2008). Testing for nested oscillation. Journal of Neuroscience Methods, 174(1), 50–61.

Rolls, E. T., Huang, C. C., Lin, C. P., Feng, J., & Joliot, M. (2020). Automated anatomical labelling atlas 3. NeuroImage, 206, 116189.

Rubinov, M., & Sporns, O. (2010). Complex network measures of brain connectivity: uses and interpretations. NeuroImage, 52(3), 1059–1069.

Schaefer, A., Kong, R., Gordon, E. M., Laumann, T. O., Zuo, X. N., Holmes, A. J., ... & Yeo, B. T. (2018). Local-global parcellation of the human cerebral cortex from intrinsic functional connectivity MRI. Cerebral Cortex, 28(9), 3095–3114.

Tadel, F., Baillet, S., Mosher, J. C., Pantazis, D., & Leahy, R. M. (2011). Brainstorm: a user-friendly

application for MEG/EEG analysis. Computational Intelligence and Neuroscience, 2011, 879716.

Tort, A. B., Komorowski, R., Eichenbaum, H., & Kopell, N. (2010). Measuring phase-amplitude coupling between neuronal oscillations of different frequencies. Journal of Neurophysiology, 104(2), 1195–1210.

Van Veen, B. D., van Drongelen, W., Yuchtman, M., & Suzuki, A. (1997). Localization of brain electrical activity via linearly constrained minimum variance spatial filtering. IEEE Transactions on Biomedical Engineering, 44(9), 867–880.

Vinck, M., Oostenveld, R., van Wingerden, M., Battaglia, F., & Pennartz, C. M. (2011). An improved index of phase-synchronization for electrophysiological data in the presence of volume-conduction, noise and sample-size bias. NeuroImage, 55(4), 1548–1565.